
%
%

\input psfig



%
%
%
%
%


\hsize=6.0truein
\vsize=8.5truein
\voffset=0.25truein
\hoffset=0.1875truein
\tolerance=1000
\hyphenpenalty=500
\def\monthintext{\ifcase\month\or January\or February\or
   March\or April\or May\or June\or July\or August\or
   September\or October\or November\or December\fi}


\font\tenrm=cmr10 scaled \magstep1   \font\tenbf=cmbx10 scaled \magstep1
\font\sevenrm=cmr7 scaled \magstep1  
\font\fiverm=cmr5 scaled \magstep1   

\font\teni=cmmi10 scaled \magstep1   \font\tensy=cmsy10 scaled \magstep1
\font\seveni=cmmi7 scaled \magstep1  \font\sevensy=cmsy7 scaled \magstep1
\font\fivei=cmmi5 scaled \magstep1   \font\fivesy=cmsy5 scaled \magstep1

\font\tentt=cmtt10 scaled \magstep1
\font\tenit=cmti10 scaled \magstep1
\font\tensl=cmsl10 scaled \magstep1

\def\twelvepoint{\def\rm{\fam0\tenrm}
   \textfont0=\tenrm \scriptfont0=\sevenrm \scriptscriptfont0=\fiverm
   \textfont1=\teni  \scriptfont1=\seveni  \scriptscriptfont1=\fivei
   \textfont2=\tensy \scriptfont2=\sevensy \scriptscriptfont2=\fivesy
   \textfont\itfam=\tenit \def\it{\fam\itfam\tenit}
   \textfont\ttfam=\tentt \def\tt{\fam\ttfam\tentt}
   \textfont\bffam=\tenbf \def\bf{\fam\bffam\tenbf}
   \textfont\slfam=\tensl \def\sl{\fam\slfam\tensl} \rm
   \hfuzz=1pt\vfuzz=1pt
   \setbox\strutbox=\hbox{\vrule height 10.2pt depth 4.2pt width 0pt}
   \parindent=24pt\parskip=1.2pt plus 1.2pt
   \topskip=12pt\maxdepth=4.8pt\jot=3.6pt
   \normalbaselineskip=14.4pt\normallineskip=1.2pt
   \normallineskiplimit=0pt\normalbaselines
   \abovedisplayskip=13pt plus 3.6pt minus 5.8pt
   \belowdisplayskip=13pt plus 3.6pt minus 5.8pt
   \abovedisplayshortskip=-1.4pt plus 3.6pt
   \belowdisplayshortskip=13pt plus 3.6pt minus 3.6pt
   \topskip=12pt \splittopskip=12pt
   \scriptspace=0.6pt\nulldelimiterspace=1.44pt\delimitershortfall=6pt
   \thinmuskip=3.6mu\medmuskip=3.6mu plus 1.2mu minus 1.2mu
   \thickmuskip=4mu plus 2mu minus 1mu
   \smallskipamount=3.6pt plus 1.2pt minus 1.2pt
   \medskipamount=7.2pt plus 2.4pt minus 2.4pt
   \bigskipamount=14.4pt plus 4.8pt minus 4.8pt}

\twelvepoint



\font\titlerm=cmr10 scaled \magstep3
\font\titlerms=cmr10 scaled \magstep1 
\font\titlei=cmmi10 scaled \magstep3  
\font\titleis=cmmi10 scaled \magstep1 
\font\titlesy=cmsy10 scaled \magstep3 	
\font\titlesys=cmsy10 scaled \magstep1  
\font\titleit=cmti10 scaled \magstep3	
\skewchar\titlei='177 \skewchar\titleis='177 
\skewchar\titlesy='60 \skewchar\titlesys='60 

\def\titlefont{\def\rm{\fam0\titlerm}
   \textfont0=\titlerm \scriptfont0=\titlerms 
   \textfont1=\titlei  \scriptfont1=\titleis  
   \textfont2=\titlesy \scriptfont2=\titlesys 
   \textfont\itfam=\titleit \def\it{\fam\itfam\titleit} \rm}


\def\preprint#1{\baselineskip=19pt plus 0.2pt minus 0.2pt \pageno=0
   \begingroup
   \nopagenumbers\parindent=0pt\baselineskip=14.4pt\rightline{#1}}
\def\title#1{
   \vskip 0.9in plus 0.45in
   \centerline{\titlefont #1}}
\def\secondtitle#1{}
\def\author#1#2#3{\vskip 0.9in plus 0.45in
   \centerline{{\bf #1}\myfoot{#2}{#3}}\vskip 0.12in plus 0.02in}
\def\secondauthor#1#2#3{}
\def\addressline#1{\centerline{#1}}
\def\abstract{\vskip 0.7in plus 0.35in
	\centerline{\bf Abstract}
	\smallskip}
\def\finishtitlepage#1{\vskip 0.8in plus 0.4in
   \leftline{#1}\supereject\endgroup}

\def\date#1{\finishtitlepage{#1}}

\def\nolabels{\def\eqnlabel##1{}\def\eqlabel##1{}\def\figlabel##1{}%
	\def\reflabel##1{}}
\def\writelabels{\def\eqnlabel##1{%
	{\escapechar=` \hfill\rlap{\hskip.11in\string##1}}}%
	\def\eqlabel##1{{\escapechar=` \rlap{\hskip.11in\string##1}}}%
	\def\figlabel##1{\noexpand\llap{\string\string\string##1\hskip.66in}}%
	\def\reflabel##1{\noexpand\llap{\string\string\string##1\hskip.37in}}}
\nolabels


\global\newcount\secno \global\secno=0
\global\newcount\meqno \global\meqno=1
\global\newcount\subsecno \global\subsecno=0

\font\secfont=cmbx12 scaled\magstep1

\def\section#1{\global\advance\secno by1
   \xdef\secsym{\the\secno.}
   \global\subsecno=0
   \global\meqno=1\bigbreak\medskip
   \noindent{\secfont\the\secno. #1}\par\nobreak\smallskip\nobreak\noindent}

\def\subsection#1{\global\advance\subsecno by1
\medskip
\noindent
{\bf\the\secno.\the\subsecno\ #1}
\par\medskip\nobreak\noindent}

\def\newsec#1{\global\advance\secno by1
   \xdef\secsym{\the\secno.}
   \global\meqno=1\bigbreak\medskip
   \noindent{\bf\the\secno. #1}\par\nobreak\smallskip\nobreak\noindent}
\xdef\secsym{}

\def\appendix#1#2{\global\meqno=1\xdef\secsym{\hbox{#1.}}\bigbreak\medskip
\noindent{\bf Appendix #1. #2}\par\nobreak\smallskip\nobreak\noindent}

\def\acknowledgements{\bigbreak\medskip\centerline{\bf
   Acknowledgements}\par\nobreak\smallskip\nobreak\noindent}


\def\eqnn#1{\xdef #1{(\secsym\the\meqno)}%
	\global\advance\meqno by1\eqnlabel#1}
\def\eqna#1{\xdef #1##1{\hbox{$(\secsym\the\meqno##1)$}}%
	\global\advance\meqno by1\eqnlabel{#1$\{\}$}}
\def\eqn#1#2{\xdef #1{(\secsym\the\meqno)}\global\advance\meqno by1%
	$$#2\eqno#1\eqlabel#1$$}


\def\myfoot#1#2{{\baselineskip=14.4pt plus 0.3pt\footnote{#1}{#2}}}
\global\newcount\ftno \global\ftno=1
\def\foot#1{{\baselineskip=14.4pt plus 0.3pt\footnote{$^{\the\ftno}$}{#1}}%
	\global\advance\ftno by1}


\global\newcount\refno \global\refno=1
\newwrite\rfile

\def\ref{[\the\refno]\nref}
\def\nref#1{\xdef#1{[\the\refno]}\ifnum\refno=1\immediate
	\openout\rfile=refs.tmp\fi\global\advance\refno by1\chardef\wfile=\rfile
	\immediate\write\rfile{\noexpand\item{#1\ }\reflabel{#1}\pctsign}\findarg}
\def\findarg#1#{\begingroup\obeylines\newlinechar=`\^^M\passarg}
	{\obeylines\gdef\passarg#1{\writeline\relax #1^^M\hbox{}^^M}%
	\gdef\writeline#1^^M{\expandafter\toks0\expandafter{\striprelax #1}%
	\edef\next{\the\toks0}\ifx\next\null\let\next=\endgroup\else\ifx\next\empty%

\else\immediate\write\wfile{\the\toks0}\fi\let\next=\writeline\fi\next\relax}}
	{\catcode`\%=12\xdef\pctsign{
\def\striprelax#1{}

\def\semi{;\hfil\break}
\def\addref#1{\immediate\write\rfile{\noexpand\item{}#1}} 

\def\listrefs{\vfill\eject\immediate\closeout\rfile
   {{\secfont References}}\bigskip{\frenchspacing%
   \catcode`\@=11\escapechar=` %
   \input refs.tmp\vfill\eject}\nonfrenchspacing}

\def\startrefs#1{\immediate\openout\rfile=refs.tmp\refno=#1}


\global\newcount\figno \global\figno=1
\newwrite\ffile
\def\fig{\the\figno\nfig}
\def\nfig#1{\xdef#1{\the\figno}\ifnum\figno=1\immediate
	\openout\ffile=figs.tmp\fi\global\advance\figno by1\chardef\wfile=\ffile
	\immediate\write\ffile{\medskip\noexpand\item{Fig.\ #1:\ }%
	\figlabel{#1}\pctsign}\findarg}

\def\listfigs{\vfill\eject\immediate\closeout\ffile{\parindent48pt
	\baselineskip16.8pt{{\secfont Figure Captions}}\medskip
	\escapechar=` \input figs.tmp\vfill\eject}}

\def\noblackbox{\overfullrule=0pt}
\def\inv{^{\raise.18ex\hbox{${\scriptscriptstyle -}$}\kern-.06em 1}}
\def\dup{^{\vphantom{1}}}
\def\Dsl{\,\raise.18ex\hbox{/}\mkern-16.2mu D} 
\def\dsl{\raise.18ex\hbox{/}\kern-.68em\partial}
\def\slash#1{\raise.18ex\hbox{/}\kern-.68em #1}
\def\lspace{}
\def\lbspace{}
\def\boxeqn#1{\vcenter{\vbox{\hrule\hbox{\vrule\kern3.6pt\vbox{\kern3.6pt
	\hbox{${\displaystyle #1}$}\kern3.6pt}\kern3.6pt\vrule}\hrule}}}
\def\mbox#1#2{\vcenter{\hrule \hbox{\vrule height#2.4in
	\kern#1.2in \vrule} \hrule}}  
\def\bar{\overline}
\def\e#1{{\rm e}^{\textstyle#1}}
\def\del{\partial}
\def\curly#1{{\hbox{{$\cal #1$}}}}
\def\curlyD{\hbox{{$\cal D$}}}
\def\curlyL{\hbox{{$\cal L$}}}
\def\vev#1{\langle #1 \rangle}
\def\psibar{\overline\psi}
\def\lform{\hbox{$\sqcup$}\llap{\hbox{$\sqcap$}}}
\def\darr#1{\raise1.8ex\hbox{$\leftrightarrow$}\mkern-19.8mu #1}
\def\half{{\textstyle{1\over2}}} 
\def\roughly#1{\ \lower1.5ex\hbox{$\sim$}\mkern-22.8mu #1\,}
\def\MSbar{$\bar{{\rm MS}}$}
\hyphenation{di-men-sion di-men-sion-al di-men-sion-al-ly}

\parindent=0pt
\parskip=5pt


\preprint{
\vbox{
\rightline{CERN-TH.7203/94}
\vskip2pt\rightline{SHEP 93/94-16}
}
}
\vskip -1cm

\title{Derivative Expansion of the Exact Renormalization Group}
\vskip -1cm
\author{\bf Tim R. Morris}{}{}
\vskip 0.5cm
\addressline{\it CERN TH-Division}
\addressline{\it CH-1211 Geneva 23}
\addressline{\it Switzerland\myfoot{$^*$}{\rm On Leave from Southampton
University, U.K.}}
\addressline{\it }
\vskip -1cm

\abstract 
The functional flow equations for the Legendre effective action, with
respect to changes in a smooth cutoff, are approximated by a derivative
expansion; no other approximation is made. This results in a set of
coupled non-linear differential equations. The corresponding differential
equations for a fixed point action have at most a countable number of
solutions that are well defined for all values of the field. We apply the
technique to the fixed points of one-component real scalar field theory in
three dimensions. Only two non-singular solutions are found: the gaussian fixed
point
and an approximation to the Wilson fixed point. The latter is used to compute
critical exponents, by carrying the approximation to second order.
The results appear to converge rapidly.
\vskip -1cm
\date{\vbox{
{CERN-TH.7203/94}
\vskip2pt{SHEP 93/94-16}
\vskip2pt{hep-ph/9403340}
\vskip2pt{March, 1994.}
}
}
\def\nonp{non-perturbative}
\def\phi{\varphi}
\def\epsilon{\varepsilon}
\def\p{{\bf p}}
\def\P{{\bf P}}
\def\q{{\bf q}}
\def\r{{\bf r}}
\def\x{{\bf x}}
\def\y{{\bf y}}
\def\tr{{\rm tr}}
\def\D{{\cal D}}
\def\ins#1#2#3{\hskip #1cm \hbox{#3}\hskip #2cm}
\def\frac#1#2{{#1\over#2}}
\nref\Wil{K. Wilson and J. Kogut, Phys. Rep. 12C (1974) 75.}
\nref\erg{Tim R. Morris, CERN / Southampton preprint CERN-TH.6977/93, SHEP
92/93-27, hep-ph/9308265, to be published in Int. J. Mod. Phys. A.}
\nref\trunc{``On Truncations of the Exact
Renormalization Group'', T.R. Morris, CERN / \hfil\break
Southampton preprint in preparation.}
\nref\truncm{``Momentum Scale Expansion of Sharp Cutoff Flow
Equations'', T.R. Morris, CERN / Southampton preprint in preparation. }
\nref\alsotrunc{A. Margaritis, G. \'Odor and A. Patk\'os, Z. Phys. C39 (1988)
109;\
 P.E. Haagensen, Y. Kubyshin, J.I. Latorre and  E. Moreno, Barcelona
preprint UB-ECM-PF\#93-20.}
\nref\suctrunc{Expansion around the semi-classical minimum of the
potential seems spectacularly to improve the convergence problem:\
N. Tetradis, C. Wetterich, preprint DESY-93-094;\
M. Alford, Cornell preprint CLNS 94/1279, hep-ph/9403324.}
\nref\hashas{A. Hasenfratz and P. Hasenfratz, Nucl. Phys. B270 (1986) 685.}
\nref\others{For applications see e.g.
U. Ellwanger and C. Wetterich Heidelberg preprint HD-THEP-94-1;\
T.E. Clark et al, Purdue preprint PURD-TH-94-01.}
\nref\formal{Formal developments are covered in  the second paper of
ref.\alsotrunc\ and \
R.D. Ball and R.S. Thorne, Oxford  preprint OUTP-93-23-P;\
M. Bonini, M. D'Attanasio and G. Marchesini, Parma preprint UPRF-94-392.}
\nref\ellw{A possibly practical method of incorporating gauge invariance
is given in U. Ellwanger, Heidelberg preprint HD-THEP-94-2.}

Our aim is to find a reliable and accurate analytic approximation
method of general applicability to \nonp\ quantum field theory
\erg--\truncm. We believe
that such a method must be based, in practice, on the Exact
Renormalisation Group\foot{a.k.a. Wilson's
renormalisation group}\Wil\ (ERG), for reasons
explained in ref.\erg.\foot{For examples of other recent work on the ERG see
refs.\alsotrunc--\ellw.}\
We particularly consider here the \nonp\ low energy behaviour
-- the Wilson fixed point\Wil\ -- of three dimensional
one-component ($Z_2$ invariant) scalar field theory, which is also the
continuum limit of the three dimensional Ising model and physically the
universality class of critical binary fluids and liquid-vapour transitions
in classical fluids\ref\zinn{See e.g.  J. Zinn-Justin,
``Quantum Field Theory and Critical Phenomena'' (1989) Clarendon Press,
Oxford.}. We choose to study this because
of its relative simplicity but the method -- just a derivative expansion of the
Legendre effective action combined with the ERG -- is certainly, appropriately
modified\ellw, of much wider applicability to \nonp\ quantum field theory.
In this letter we concentrate on developing a practical framework for such
an expansion. The explicit calculations are only taken to second order, which
is not sufficient (with our choice of cutoff) for our estimates of exponents
to be competitive with the worlds best present estimates\zinn; nevertheless,
already the differences, between our results and those, are 6 or less
times their quoted error, and our error appears to roughly halve with each
new order of approximation. Finally we should emphasize that, in calculating
directly the effective lagrangian, we may derive much more of
physical interest than just critical exponents.

The reader may well be asking why such a natural approximation scheme, built
around ideas nearing twenty years old\Wil, has not been well studied before.
It is as well to realise however that there are many contenders for methods
of approximation, that sound perhaps equally natural, but do not work. It
is also true that the computation must be organised carefully: merely expanding
vertices in invariant polynomial combinations of momenta will soon grind
to a halt through shear complexity. We list below methods that do not work
in general.

\item{a)} A sequence of truncations of Dyson-Schwinger equations, and other
methods where renormalisability is a serious problem\erg.
\item{b)} A sequence of truncations of the ERG, because these do not converge
beyond a certain point, and because no completely reliable method could be
found to reject the many spurious solutions that are also
generated\trunc--\alsotrunc\ (see however \suctrunc).
\item{c)} Direct numerical approaches to (truncations of) the ERG\erg.
\item{d)} Approximations other than those that can be formulated for
one particle irreducible parts of the vertices of the Wilson effective
action\erg.
\item{e)} Momentum (i.e. derivative) expansion of the ERG with smooth cutoff,
if the effective width over which the cutoff varies is any less than the cutoff
itself\erg.
\item{f)} Momentum scale expansion of the ERG with sharp cutoff because
only certain truncations are calculable in closed form\truncm\ -- but
this is not good enough, c.f. (b).

Because of (d) we use flow equations formulated directly for the Legendre
effective action with respect to changes in some {\sl infrared cutoff}
scale $\Lambda$. This gives directly the one particle irreducible parts
of the Wilson effective action as detailed in ref.\erg.\foot{ It should
be realized that {\sl all} the ERG equations are straightforwardly
equivalent to each other, up to choice of cutoff function; for example
the Wegner-Houghton equations\ref\weg{F.J. Wegner and A. Houghton,
Phys. Rev. A8 (1973) 401.}\ are the Polchinski
equations\ref\pol{J. Polchinski, Nucl. Phys. B231 (1984) 269.}\
in the limit of sharp cutoff, while these
are equivalent to Wilson's equations\Wil\ under the momentum dependent
change of variables $\phi\to\phi\sqrt{K}$ that eliminates the cutoff
function $K$\pol\ from the propagator. For other comparisons see refs.
\erg\ and \ref\weginf{F.J. Wegner, J. Phys. C7 (1974) 2098.}.}\
Because of (c) we must
approximate the momentum dependence and the only efficacious method we know
of, is a momentum expansion\erg. Because of (e) and (f) we use a smooth cutoff
with an effective width at least $\Lambda$ (that is with little variation
on scales less than $\Lambda$). This means that the cutoff will only
somewhat suppress, and not eliminate, low energy modes. Finally because of (b)
we perform only the derivative expansion and do not at all expand in the
field $\phi$.

The partition function is  defined as
\eqn\zorig{\exp W[J]=\int\!\D\phi\
\exp\{-\half\phi.C^{-1}.\phi-S_{\Lambda_0}[\phi]+J.\phi\}\ \ .}
The notation is the same as previously\erg--\truncm,
so two-point functions are often regarded as matrices in position or
momentum ($\q$)
space, one-point functions as vectors, and contractions indicated by a
dot. We work in $D$ euclidean dimensions with a single real scalar field
$\phi$. The definition differs only in
 that here we mean $S_{\Lambda_0}$
to be the {\sl full} bare action for the theory; the $C\equiv C(q,\Lambda)$
term is taken to be an `additive cutoff', for convenience. Previously we took
 a formulation with a multiplicative cutoff
$\theta_\epsilon$, so that a
cutoff massless propagator took the form $\theta_\epsilon(q,\Lambda)/q^2$.
Of course the two are trivially related, in this case
\eqn\cutrel{\theta_\epsilon(q,\Lambda)=q^2C(q,\Lambda)/[1+q^2C(q,\Lambda)]\ \
.}
The reason for the change is that the smooth cutoff will have to scale
anomalously along with $\phi$ to reach non-gaussian fixed points, and this
scaling is homogeneous for smooth additive cutoffs, but takes the form \cutrel\
for smooth multiplicative cutoffs.

For $C(q,\Lambda)$ to be an infrared cutoff it follows that $C$ should be
small for $q<\Lambda$, ideally $C$ tending to zero as $q\to0$,
and $q^2C(q,\Lambda)$ should be large for $q>\Lambda$, tending to infinity as
$q\to\infty$. The ultra-violet regularisation need not be discussed
explicitly\erg, in fact since we will concentrate exclusively on the
neighbourhood of fixed points it can be ignored entirely. From \zorig\ we
have
$${\partial\over\partial\Lambda}W[J]=
-{1\over2}\left\{ {\delta W\over
\delta J}.{\partial C^{-1} \over\partial\Lambda}.{\delta
W\over\delta J} + \tr\left({\partial C^{-1} \over\partial\Lambda}.{\delta^2
W\over\delta J\delta J}\right)\right\}\quad ,$$
which on rewriting in terms of the Legendre effective action $\Gamma$ gives
(as in ref.\erg),
$${\partial\over\partial\Lambda}\Gamma[\phi]=
-{1\over2}\tr\left[{1\over C}{\partial C\over \partial\Lambda}
.\left( 1+C.{\delta^2\Gamma\over\delta\phi\delta\phi}\right)^{-1}\right]
\ \ .$$
$\Gamma$ is defined by
$\Gamma[\phi]+\half\phi.C^{-1}.\phi=-W[J]+J.\phi$,
where now $\phi=\delta W/\delta J$ is the classical field.
 As in ref.\erg\ it is helpful to write the trace as an integral over
momentum space and factor out the $D$-dimensional solid angle:
\eqn\unscaledG{{\partial\Gamma\over\partial\Lambda}=
-{\Omega\over2}\int_0^\infty\!\!\!\!dq\  {q^{D-1}\over C(q,\Lambda)}
{\partial C(q,\Lambda)\over\partial\Lambda}
\left\langle\left[1+C.{\delta^2\Gamma\over\delta\phi\delta\phi}
\right]^{-1}\mkern-23mu(\q,-\q)\right\rangle\ \ ,}
where $\Omega=2/[\Gamma(D/2) (4\pi)^{D/2}]$ is the solid angle of a
$(D-1)$-sphere divided by $(2\pi)^D$, the brackets $\langle\cdots\rangle$
mean an average over all directions of the momentum $\q$, and we have written
the momentum dependence of the traced inverse two-point function explicitly.

At a fixed point the field scales anomalously as $\phi\sim\Lambda^{d_\phi}$
with $d_\phi=\half(D-2+\eta)$, $\eta$ being the anomalous scaling
dimension. Dimensional analysis then indicates that we must have
\eqn\scaleC{C(q,\Lambda)\to \Lambda^{\eta-2}{\tilde C}(q^2/\Lambda^2)}
for some ${\tilde C}$ as we approach the fixed point, if $\Gamma$ is to
become independent of $\Lambda$ -- i.e. fixed -- in that limit.
What is happening is that the anomalous scaling of $\phi$ induces also
an anomalous scaling of the width over which the smooth cutoff changes;
it is because this width is zero in the sharp case that anomalous scaling
may be incorporated in that case automatically\weg\truncm.
Now we rewrite the equation in terms of dimensionless
quantities\foot{From now on we drop the tilde on the scaled $C$.}\ via \scaleC,
$\q\mapsto \Lambda\q$, $\phi(\Lambda\q)\mapsto \Lambda^{d_\phi-D}\phi(\q)$,
and $t=\ln(\Lambda_0/\Lambda)$.  Finally we rescale
the fields and effective action as $\phi\mapsto\phi\sqrt{\Omega/2\zeta}$,
$\Gamma\mapsto (\Omega/2\zeta)\Gamma$ to absorb the factor $\Omega/2$
in \unscaledG. $\zeta$ is a normalization factor which will be chosen later
for convenience. The result is
\eqn\scaledG{\eqalign{
({\partial\over\partial t}&+d_\phi\Delta_\phi+\Delta_\partial
-D)\Gamma[\phi] =\cr
&-\zeta\int_0^\infty\!\!\!\!\!dq\, q^{D-1}
\left({q\over C(q^2)}{\partial C(q^2)\over\partial q} +2- \eta\right)
\left\langle\left[1+C.{\delta^2\Gamma\over\delta\phi\delta\phi}
\right]^{-1}\mkern-23mu(\q,-\q)\right\rangle\ \ .\cr}}
In here $\Delta_\phi=\phi.{\delta\over\delta\phi}$ is the `phi-ness' counting
operator: it counts the number of occurences of the field $\phi$ in a
given vertex. $\Delta_\partial$ may be expressed as
$$\Delta_\partial =D+\int\!{d^Dp\over(2\pi)^D}\,\phi(\p)p^\mu
{\partial\over\partial p^\mu} {\delta\over\delta\phi(\p)} $$
i.e. the momentum scale counting operator $+D$. Operating on a given
vertex it counts the total number of derivatives acting on the fields
$\phi$.

We write $\Gamma$ as the space-time integral
of an effective lagrangian expanded in powers of derivatives.
Each linearly independent (under integration by parts) scalar combination
of differentiated fields $\phi$ will be accompanied by its own general
($t$ dependent) coefficient function of the fields:
\eqn\gexp{\Gamma[\phi]=\int\! d^Dx\,\{
V(\phi,t)+\half(\partial_\mu\phi)^2K(\phi,t)+(\partial_\mu\phi)^4H_1(\phi,t)
+(\lform\phi)^2H_2(\phi,t)+\cdots\} \ \ .}
(We have not written all 3 four-derivative terms).
 Substituting this into \scaledG, expanding the Right Hand Side (RHS) of
\scaledG\ up to some maximum power of derivatives -- this is the
approximation -- and equating coefficients of the linearly independent
basis set, yields a set  of coupled non-linear differential equations for
the coefficient functions.

The derivative expansion of the RHS of \scaledG\ may be computed as follows.
Regard $\left[1+C.{\delta^2\Gamma\over\delta\phi\delta\phi}\right]^{-1}$
as a differential operator: \eqnn\dQ \eqnn\Q
$$\eqalignno{\left[1+C.{\delta^2\Gamma\over\delta\phi\delta\phi}
\right]^{-1}\mkern-23mu(\q,-\q) &=
\int\! d^D\!x\,d^D\!y\ \e{-i\q.\x}
\left[1+C.{\delta^2\Gamma\over\delta\phi\delta\phi}\right]^{-1}\mkern-23mu(\x,\y)
\ \e{i\q.\y}\cr
&\equiv\int\!d^D\!x\ Q &\dQ\cr
{\rm where} \hskip 2cm Q &=\e{-i\q.\x}
\left[1+C.{\delta^2\Gamma\over\delta\phi\delta\phi}\right]^{-1}\!\!
\e{i\q.\x}\ \ .&\Q\cr}$$
$Q$ is a function of $\q$, and $\phi(\x)$ and its derivatives evaluated at
$\x$.
Using \Q, we see that $Q$ satisfies the differential equation
\eqn\iter{Q={1\over 1+{\cal V}}+{1\over 1+{\cal V}}
\left\{{\cal V}Q-\e{-i\q.\x}C(-\lform)
{\delta^2\Gamma\over\delta\phi\delta\phi}\, \e{i\q.\x}Q\right\}
\ \ ,}
where $C$ and ${\delta^2\Gamma\over\delta\phi\delta\phi}$ are also regarded
as differential operators -- acting on all terms to their right. For
convenience the function ${\cal V}(\phi(\x),\q)$ is introduced and defined
to be the expression obtained by dropping from
\eqn\calV{C(q^2)\ \e{-i\q.\x}{\delta^2\Gamma\over\delta\phi\delta\phi}\,
\e{i\q.\x}}
all terms containing differentials of $\phi$.
With this definition the derivative expansion of $Q$ may be straightforwardly
performed by iteration of \iter. To complete the computation of the RHS of
\scaledG, it only remains to perform the angular average and compute the
resulting radial $q$ integrals. But the angular averages are easily translated
into invariant tensors: $\langle q^\mu q^\nu \rangle=q^2\delta^{\mu\nu}/D$
etc.

We will require that $t$-independent (fixed point) solutions for the
coefficient functions: $V(\phi),K(\phi),H_1(\phi),H_2(\phi),\dots$
be non-singular for all real
$\phi$. In addition we will require $\Gamma$ to be parametrized so that
$K(0)\ne0$.\foot{The obvious stability requirements do not need to be
separately addressed since unstable solutions are not allowed by the
equations.}\ As a result it will be seen that the differential equations
have at most a countable number of fixed point solutions (at least for the
first two orders and presumably to any order
of approximation in the derivative expansion). The coefficient functions
of linearized perturbations about these solutions -- ie. of the eigenoperator
spectrum -- will be required to grow no faster than a power of $\phi$ as
$\phi\to\infty$. It will be seen that this forces  power law growth (as
$\phi\to\infty$), the power being given by the scaling dimension, and
results in a quantized spectrum.

For a general cutoff function $C(q^2)$, for example based on an exponential,
this would appear to leave $\eta$ as an undetermined parameter however.
In this case there is presumably a range of reasonable approximations,
which can be thought of as labelled by the effective width of the cutoff
(e.g. via \cutrel\ modified by the value of $K(0)$).
 On the other hand, dimensional
analysis indicates that if $C(q^2)$ is homogeneous in $q^2$ then the flow
equation \scaledG\ is invariant under a global scaling symmetry. This
symmetry then overdetermines the equations so that solutions exist only
for certain discrete values of $\eta$. (This is similar to the case of
sharp cutoff\truncm\weg\ where the flow equations are invariant under the
phi-ness scaling symmetry $\phi\to\lambda\phi$). We pursue this form for the
cutoff
here, and therefore require $C(q^2)=q^{2k}$ for $k$ a non-negative integer.
(The proportionality constant may be set to 1 by a rescaling of $\phi$).
The scaling symmetry is characterised by the following dimensions, as
trivially follows from \gexp\ and the definition of $\Gamma$ (see above
\unscaledG):
\eqn\ssym{\eqalign{
&[\partial_\mu]=[q_\mu]=1, \quad\quad [\phi]=k+D/2,\cr
{\rm hence}\hskip 2cm [V]=&D,\quad  [K]=-2(k+1),\quad [H_i]=-D-4(k+1)
\ \ .\cr}}
(Of course these dimensions should in no way be confused with their physical
scaling dimensions).

Using \cutrel\ one finds $[1-\theta_\epsilon(q,\Lambda)]/q^2=1/[q^2(1+
q^{2k+2})]$. This is a massless propagator with effective
U.V. cutoff as it appears for the Wilson action\erg. One sees that this choice
of cutoff corresponds to a form of Pauli-Villars or higher derivative
regularisation.  It may be that other cutoffs give stronger convergence,
so the more general case deserves its own investigation.
{}From our general arguments\erg\ we expect that the momentum expansion
has slower convergence the higher the value of $k$. On the other hand we
require $k>D/2-1$ if the momentum integral in \scaledG\ is to be
U.V. convergent to all orders in the momentum expansion. Therefore
we choose $k$ to be the least integer larger than $D/2-1$.
The cutoff function is now completely
determined. The inequality $k>D/2-1$ may be derived by
thinking graphically about the RHS of \scaledG: At a given order $p^{2m}$
($m>1$) in the momentum expansion one obtains an expansion in one loop
graphs with $N=1,2,\cdots$ vertices that behave as $\sim q^{2m}$ as
$q\to\infty$, $N+1$ propagators that behave as $\sim 1/q^{2m}$ as $q\to\infty$,
and an insertion $\sim 1/C(q)$. Putting these together with the measure
factor $\sim q^D$ one finds the stated inequality. (Actually one should
consider
also the $m=0$ case for which propagators go as $\sim1/q^2$ and vertices are
constant, but this yields a weaker constraint.)

We are finally ready to consider the explicit example.  Thus from now on
we set $D=3$ and $k=1$. To lowest order we drop {\sl all} the derivative
terms ($\partial_\mu\phi$ etc.) from the RHS of \scaledG. This means that the
coefficient functions $K,H_1,H_2,\dots$ satisfy linear equations given by the
vanishing of the LHS of \scaledG. This implies that at fixed points
$K(\phi)\propto\phi^{-2\eta/1+\eta}$,
while the requirement that $K(\phi)$ be non-singular and $K(0)\ne0$ implies
we must have $\eta=0$ (as expected at this order) and $K(\phi)$ a
constant, which we set to 1 by \ssym. Similarly we determine $H_i(\phi),
\cdots\equiv0$.
$Q$ is just $1/(1+{\cal V})$ to this order, and by \calV, ${\cal V}=
q^2[q^2+V''(\phi)]$. (We use primes to denote differentiation w.r.t. $\phi$).
Substituting \dQ\ into \scaledG\ and performing the $q$ integral we find
\eqn\first{{\partial\over\partial t}V(\phi,t) +\half\phi V'(\phi,t)-3V(\phi,t)
=-1\bigg/\sqrt{2+V''(\phi,t)}\ \ .}
Here, and from now on, we fix $\zeta=1/2\pi$. Solutions to this equation are
qualitatively very similar to those of the
equivalent sharp cutoff equation\hashas\trunc. For any putative non-trivial
fixed point solution such that $V(\phi,t)\equiv V(\phi)$, some analysis
shows that either it exists for all real $\phi$, in which case it behaves,
for $\phi\to\infty$, as
\eqn\firsty{V(\phi)=A\phi^6+{1\over4\sqrt{15A}\phi^2}+O(1/\phi^6)}
for some positive constant $A$ (an isolated one parameter set, as follows
from the analysis of (15)), or else it is only defined for $|\phi|<
\phi_c$ because the solution ends at a singularity of the form
$$V(\phi)\sim \biggl({9\over4\phi_c}\biggr)^{2/3}
(\phi_c-\phi)^{4/3}\hskip2cm{\rm for}
\quad \phi\approx+\phi_c\ \ .$$
(We have not shown the lower order singular behaviour and regular part.
Note that neither here nor at $O(\partial^2)$ does $V(\phi)$ {\sl diverge}
at the singularity.)
There is also an analogous ``slow rollover'' behaviour\trunc).
With $\partial V/\partial t\equiv 0$, \first\ is a second order
ordinary differential equation. Thus at a fixed point, solutions are
generically labelled by two parameters; however there are two conditions to be
met: that $V(\phi)$ have the behaviour \firsty\ and that $V(\phi)$ be
symmetric under $\phi\leftrightarrow-\phi$, equivalently $V'(0)=0$. (
In a non-symmetric theory this condition is replaced by requiring \firsty\
also for $\phi\to-\infty$ with a possibly different constant $A$).
Thus there are
at most a discrete set of acceptable fixed point solutions to \first. We
have searched numerically for these and find only two: the trivial gaussian
solution $V(\phi)=1/\sqrt{18}$ and the solution ``$\partial^0$'' displayed in
fig.\fig\Vs{ }, which we expect to be an approximation to the
Wilson fixed point. It may be characterised by \first\ and the ``effective
mass'' $\sigma=V''(0)=-.5346$, negative as expected\trunc.\foot{All
numerical values reported in this
letter have been determined to an accuracy greater than the number of
significant figures displayed.}\
For small perturbations about this
solution we write $V(\phi,t)=V(\phi)+\delta V(\phi,t)$, with,
by separation of variables,
\eqn\evec{\delta V(\phi,t)\propto v(\phi)\, \e{\lambda t}\ \ .}
Substituting in \first\ one obtains \nfig\Ks{}\
\eqn\firstv{\half\phi v'(\phi)+(\lambda-3)v(\phi)=
\half v''(\phi)[2+V''(\phi)]^{-3/2}\ \ .}
Again we expect solutions labelled by two parameters, however in this case
by linearity we can choose $v(0)=1$, and by symmetry we require $v'(0)=0$.
Thus solutions $v(\phi)$ are unique, given $\lambda$. However for large
$\phi$, $v(\phi)$ is generically a superposition
of $\sim\phi^{6-2\lambda}$, which is that expected from \evec\ and scaling
arguments, and $\sim\exp({1\over8}[30A]^{3/2}\phi^8)$. Requiring zero
coefficient for the latter restricts the allowed values of $\lambda$ to a
discrete set. We found just one positive eigenvalue,
which yields the correlation length
critical exponent\Wil\ through $\nu=1/\lambda$, and determined the exponent
of the first correction to scaling $\omega=-\lambda$, where $\lambda$ is the
 least negative eigenvalue.  The corresponding solutions $v(\phi)$ are
displayed in figs.\fig\nusomegas{ }.
 Our results are displayed in Table 1. These can be
compared to the equivalent sharp cutoff\hashas\trunc\ results $\nu=.6895$,
$\omega=.5952$. 
(It is worth remarking that with these methods one
very readily improves on the results for exponents given in these refs.)

Working to $O(\partial^2)$ in the derivative expansion we determine, as before,
that coefficient functions $H_i(\phi),\cdots$ vanish identically. To lowest
order $Q$ is $1/(1+{\cal V})$ with ${\cal V}=q^2[q^2 K(\phi)+V''(\phi)]$,
but now we must iterate \iter\ to $O(\partial^2)$ using $C(-\lform)=-\lform$
and ${\delta^2\Gamma\over\delta\phi\delta\phi}=V''
-K'\partial_\mu\phi\partial_\mu -\half(\partial_\mu\phi)^2K''-(\lform\phi)K'
-K\lform$. After a long but straightforward computation, involving also
integration by parts of \dQ\ and the angular average, and performing the
$q$ integrals, the RHS of \scaledG\ may be cast in the form \gexp. Comparing
both sides of \scaledG\ we obtain \eqnn\secondV\eqnn\secondK\
$$\eqalignno{&\phantom{\hbox{and}\hskip 1cm}
{\partial V\over\partial t}+{1\over2}(1+\eta)\phi V'-3V=
-{1-\eta/4\over\sqrt{K}\sqrt{V''+2\sqrt{K}} } &\secondV\cr
&\hbox{and}\hskip 1cm
{\partial K\over \partial t}+{1\over2}(1+\eta)\phi K' +\eta K=
\left(1-{\eta\over4}\right)\Biggl\{ {1\over48}{24KK''-19(K')^2\over
K^{3/2}(V''+2\sqrt{K})^{3/2}} &\secondK\cr
&-{1\over48}{58V'''K'\sqrt{K}+57(K')^2+(V''')^2K\over K(V''+2\sqrt{K})^{5/2}}
+{5\over12}{(V''')^2K+2V'''K'\sqrt{K}+(K')^2\over\sqrt{K}(V''+2\sqrt{K})^{7/2}}
\Biggr\}\ \ .\cr}$$
Considering again the fixed point equations (with $\partial V/\partial
t\equiv0$,
$\partial K/\partial t\equiv0$) we see that, by substitution of \secondV\ into
\secondK, it is possible to cast these equations in the form of two
simultaneous
second order ordinary differential equations. A little thought shows that this
generalises: at any order of a derivative expansion, substitution of the
lower order equations in the higher order equations will reduce them to
simultaneous second order differential equations, arising ultimately as a
consequence of the term ${\delta^2\Gamma\over\delta\phi\delta\phi}$ in
\scaledG. There are now two ways the solutions can end at singularities,\foot{
$V(\phi)\sim {\rm const.} (\phi_c-\phi)^m$, $K(\phi)\sim {\rm const.}
 (\phi_c-\phi)^n$ with  $(m,n)=(4/3,2/3)$ and $(8/5,-4/5)$.}\ but
we find the essential point remains the same: if $V(\phi)$ and $K(\phi)$
exist for all real $\phi$, and are not constant,
 then again they must behave, for $\phi\to\infty$, to
leading order according to their scaling dimension: \eqnn\Vy\eqnn\Ky\
$$\eqalignno{V(\phi) &= A_V \,\phi^{6/1+\eta}+
{1+\eta\over4\sqrt{6(5-\eta)A_KA_V}}\,\phi^{-2(1-\eta)/1+\eta}
+O\left(\phi^{-(6-2\eta)/1+\eta}\right) &\Vy\cr
K(\phi) &= A_K\, \phi^{-2\eta/1+\eta}+{1\over6}
{(1-\eta/2)^2\over(1+\eta)\sqrt{6(5-\eta)A_V}} \,\phi^{-(4+\eta)/1+\eta}
+O\left(\phi^{-8/1+\eta}\right) &\Ky\cr}$$
for some positive constants $A_V,A_K$. (Now an isolated two parameter set;
this then accounts for the four
behaviours expected. We have also investigated \secondV\ and \secondK\
numerically, and have checked analytically that \Vy\ and \Ky\ are the only
possible {\sl power law} behaviours (as $\phi\to\infty$) excepting several
special possibilities which however overdetermine the five parameters).

Since $\eta$ is so far a free parameter,
solutions are generically labelled by five parameters. However there are now
also five conditions to be met -- which may be given as \Vy,\Ky, $V'(0)=0$,
$K'(0)=0$, and $K(0)=1$. This last condition can be chosen using the scaling
symmetry \ssym. Thus for these ``non-linear eigenvalue equations'' there are
at most a discrete set of allowed values for $\eta$ and associated solutions
$V(\phi)$, $K(\phi)$. We have searched numerically for these and again find
only two: the gaussian solution ($\eta=0$, $K(\phi)=1$, $V(\phi)=1/\sqrt{18}$),
and an approximation to the Wilson fixed point displayed in figs.\Vs\ and
\Ks.  It is characterised by \secondV, \secondK, the solution $\eta$ given in
Table 1, and $\sigma=V''(0)=-.3782$.

As before, we linearize about this fixed point solution: $\delta V(\phi,t)=
\epsilon\, v(\phi)\, \e{\lambda t}$, $\delta K(\phi,t)=
\epsilon\, k(\phi)\, \e{\lambda t}$, $\epsilon$ infinitessimal. Linearity and
symmetry allow us to choose $v(0)=1$, $v'(0)=0$ and $k'(0)=0$, while
requiring that $v$ and $k$ behave for large $\phi$ according to their
scaling dimension as $\phi^{2(3-\lambda)/1+\eta}$ and
$\phi^{-2(\lambda+\eta)/1+\eta}$  respectively, and not as growing
(and oscillating\foot{ $\sim\exp\{ c(1\pm i\sqrt{3})\phi^p\}$, where
$p={7\over3}{4-\eta\over1+\eta}$
and $c>0$ is a calculable constant.}\
) exponentials of powers of $\phi$,
once again turns these equations into eigenvalue
equations for $\lambda$. As before we find just one positive eigenvalue and
determine the first negative eigenvalue. These yield the values for $\nu$
and $\omega$ given in Table 1 below. The corresponding solutions for $v$
and $k$ are given in figs. \nusomegas\ and \fig\ks{ }.
$$\vbox{\offinterlineskip\hrule\halign{\vrule#&&\strut\ #\ \hfil\vrule\cr
&\hfil Approx'&\hfil$\eta$&\hfil$\nu$&\hfil$\omega$\cr
\noalign{\hrule}
&\hfil $O(\partial^0)$&\hfil0&.6604&.6285\cr
&\hfil $O(\partial^2)$&.05393&.6181&.8975\cr
&\hfil Worlds Best&.035(3)&.631(2)&.80(4)\cr}\hrule}$$
Table 1. Results from the first two orders of the derivative expansion
compared to the worlds best determinations\zinn. (The
latter are combined results from the $\epsilon$ expansion, summed perturbation
theory and lattice methods).\foot{A recent
lattice Monte Carlo RG study however
gives a much lower $\eta$, and $\omega$ and
$\nu$ much closer to our $O(\partial^2)$ results\ref\baillie{
C.F. Baillie et al, Los Alamos preprint LA-UR-91-2853.}.}\
All other exponents pertinent to the fixed point
itself are related to $\eta$ and $\nu$ by scaling relations.

However, we also find solutions $v$ and $k$ with eigenvalue $\lambda
=0$. This is a simple consequence of the scaling symmetry \ssym. These
are not physical but rather are components of a redundant operator\weginf\
corresponding to a reparametrization of the effective action:
In general a change of variables $\phi_\x={\tilde\phi}_\x+
\epsilon\,\Phi_\x[{\tilde\phi}]$
in \zorig\ (with $J.\phi$ replaced by $J.{\tilde\phi}$)
induces a change in the effective action of
$\delta\Gamma=F.{\delta\Gamma\over\delta\phi}$ with
$F_\x[\phi]=\epsilon\,\exp(-W[J])\Phi_\x[\delta/\delta J]\exp(W[J])$
and a change to the definition of the cutoff functional\weginf, in effect for
the classical field $\delta\phi_\x=F_\x$. A general choice of $F$ that
leaves \gexp\ invariant in form at this order is $F_\x[\phi]\propto
\{f\bigl(\phi(\x)\bigr)+\alpha x^\mu\partial_\mu\phi(\x)\}$
for $f$ any function and $\alpha$ any constant. A redundant operator
corresponding to this $F$ must satisfy $(v,k)\propto(fV'-3\alpha V\,
,\,fK'+2f'K-\alpha K)$ for some choice of $f$ and $\alpha$.
Such is the case for the solutions at $\lambda=0$,
with $f(\phi)=5\alpha\phi/2$ and $\alpha\ne0$.

\acknowledgements

It is a pleasure to thank Simon Catterall, Poul Damgaard, Dan Freedman,
Tim Hollowood, and Ulli Wolff for their interest and discussions.

\listrefs
\midinsert
\centerline{
\psfig{figure=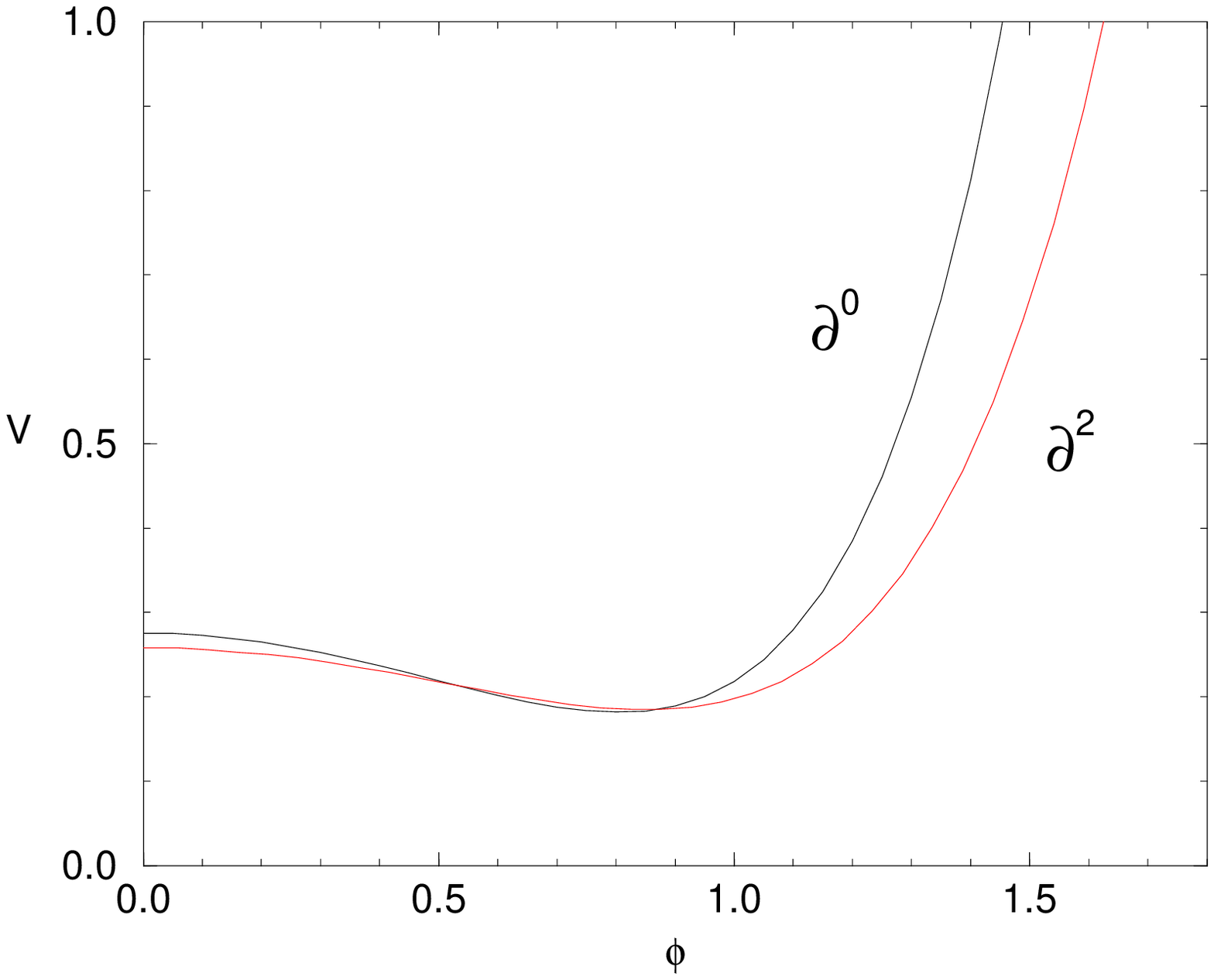,width=4in}}
\vskip 0in
\centerline{\vbox{{\bf Fig.1.} Solutions for $V(\phi)$ at order $\partial^0$
and $\partial^2$ in the derivative expansion.
}}
\endinsert
\midinsert
\centerline{
\psfig{figure=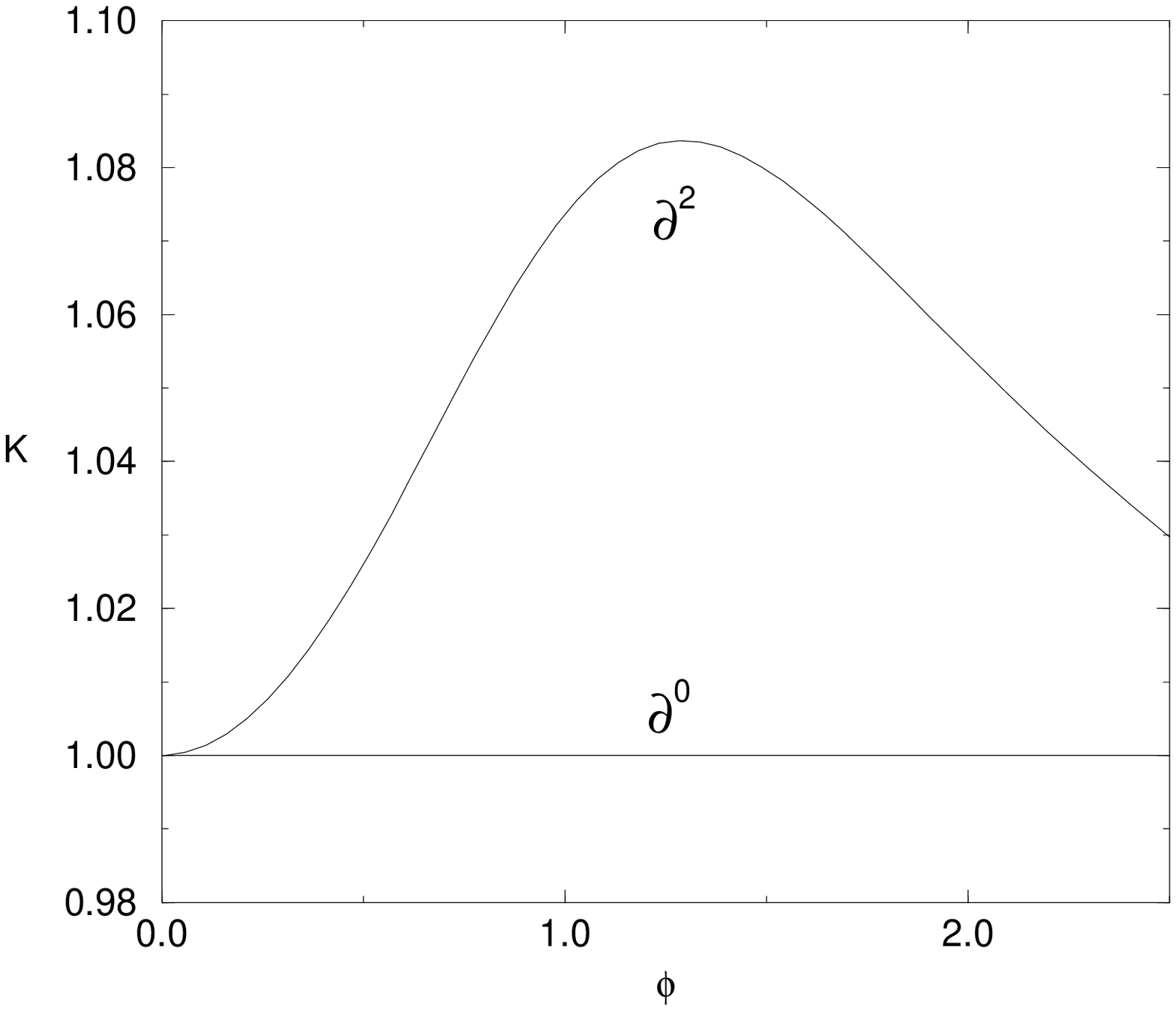,width=4in}}
\vskip 0in
\centerline{\vbox{{\bf Fig.2.} Solutions for $K(\phi)$ at order $\partial^0$
and $\partial^2$ in the derivative expansion.
}}
\endinsert
\midinsert
\centerline{
\psfig{figure=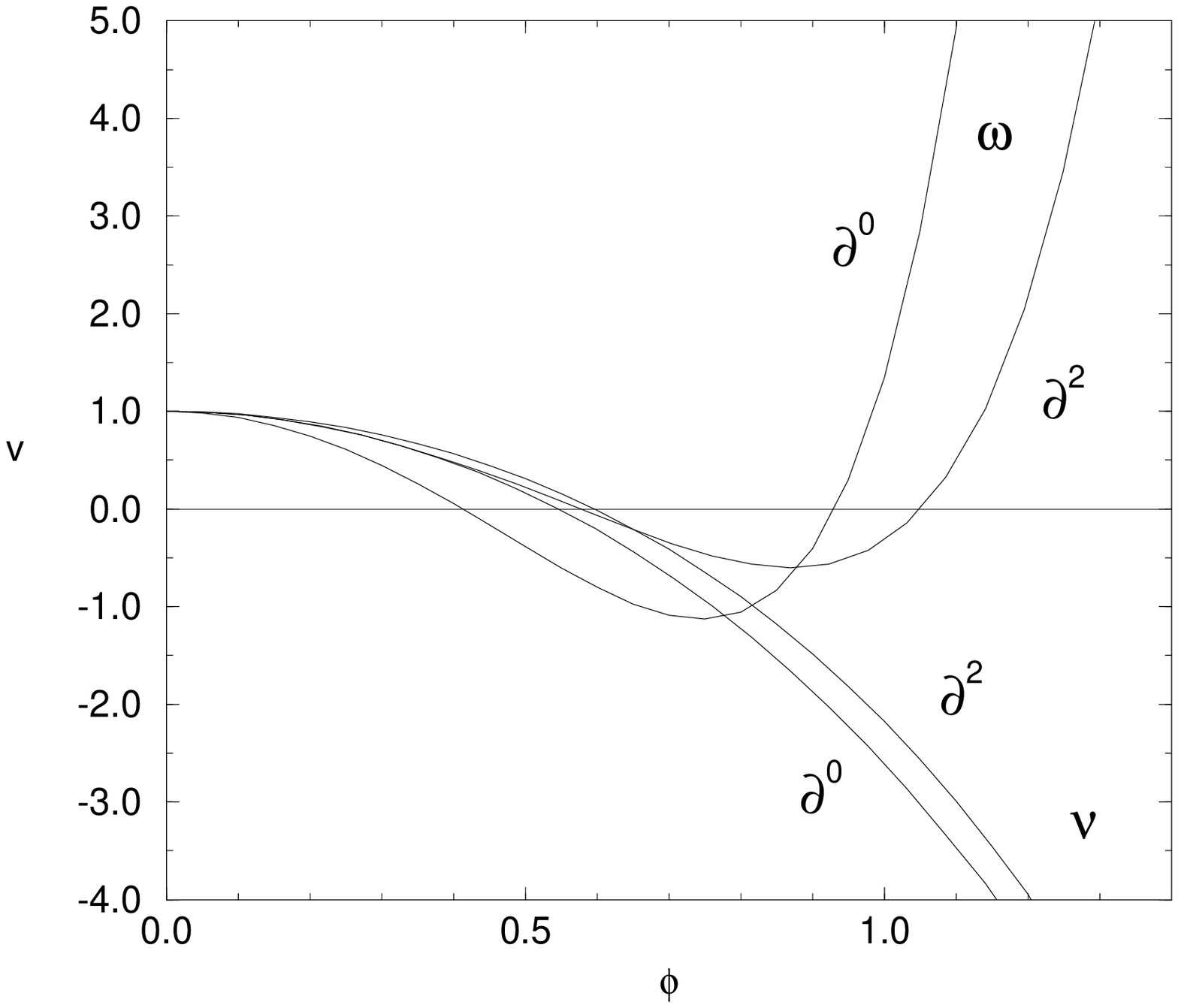,width=4in}}
\vskip 0in
\centerline{\vbox{{\bf Fig.3.} The $v(\phi)$ component
of the relevant scaling operator (associated with $\nu$), and the least
irrelevant scaling operator (associated with $\omega$),  at order $\partial^0$
and $\partial^2$.
}}
\endinsert
\midinsert
\centerline{
\psfig{figure=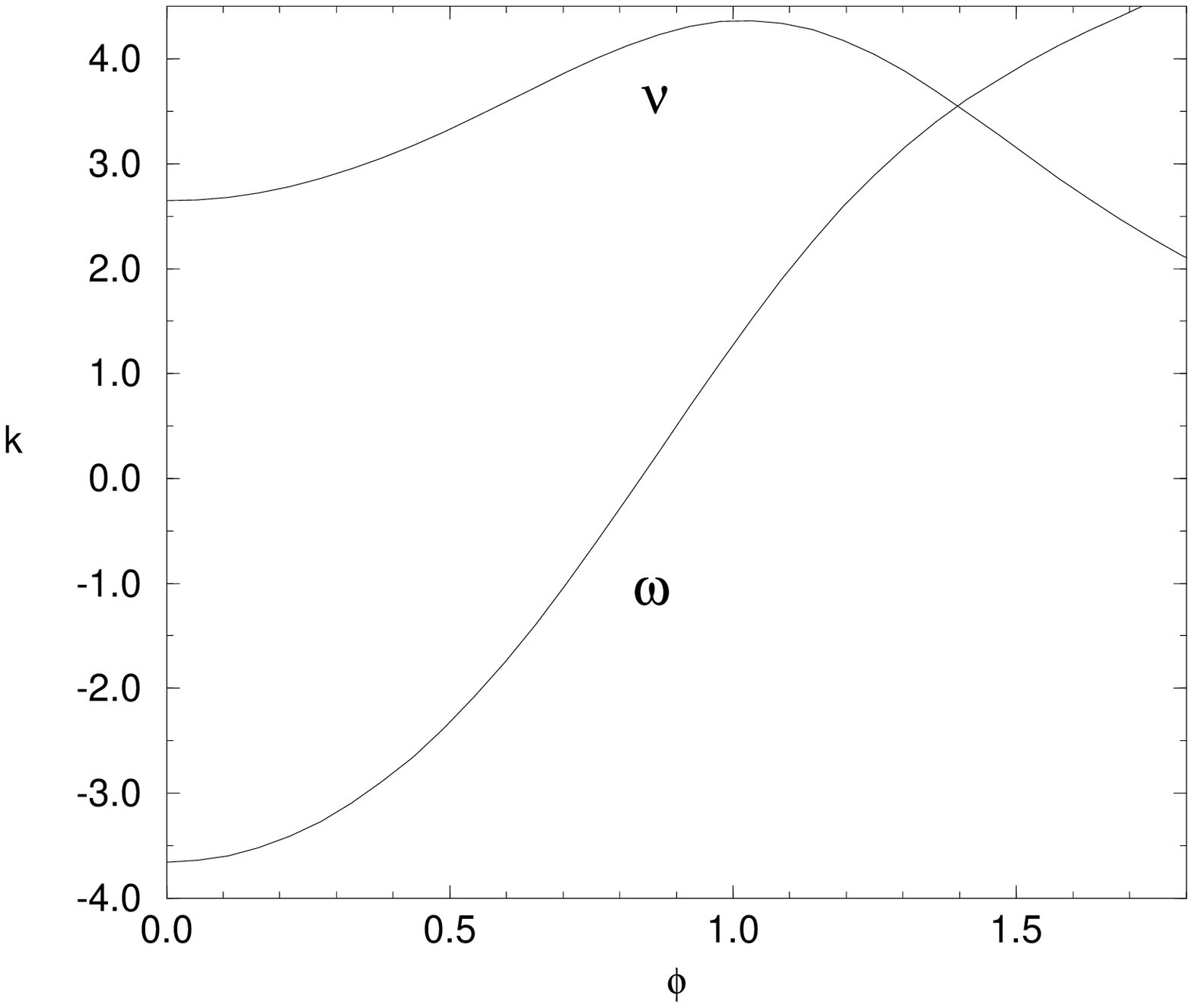,width=4in}}
\vskip 0in
\centerline{\vbox{{\bf Fig.4.} The $k(\phi)$ component
of the relevant scaling operator (associated with $\nu$),
multiplied by a factor of 10 (for display
purposes), and the least
irrelevant scaling operator (associated with $\omega$), at order  $\partial^2$.
}}
\endinsert

\end